# ENSEMBLE REGRESSION MODELS FOR SOFTWARE DEVELOPMENT EFFORT ESTIMATION: A COMPARATIVE STUDY

Halcyon D. P. Carvalho, Marília N. C. A. Lima, Wylliams B. Santos
and Roberta A. de A.Fagunde

Department of Computer Engineering, University of Pernambuco, Brazil

**ABSTRACT**

*As demand for computer software continually increases, software scope and complexity become higher than ever. The software industry is in real need of accurate estimates of the project under development. Software development effort estimation is one of the main processes in software project management. However, overestimation and underestimation may cause the software industry loses. This study determines which technique has better effort prediction accuracy and propose combined techniques that could provide better estimates. Eight different ensemble models to estimate effort with Ensemble Models were compared with each other base on the predictive accuracy on the Mean Absolute Residual (MAR) criterion and statistical tests. The results have indicated that the proposed ensemble models, besides delivering high efficiency in contrast to its counterparts, and produces the best responses for software project effort estimation. Therefore, the proposed ensemble models in this study will help the project managers working with development quality software.*

## 1. INTRODUCTION

Software Engineering is a computing branch focused on the specification, development, maintenance of software using well-defined principles, methods, procedures, and project management practices [1].

Software project management has a wider scope than the software engineering process as it involves stakeholders management, communication, which are performed to achieve a predefined product, result, or service. A critical issue in the software project management process is the estimated effort, resources, cost, and time spent in software development lifecycle [2].

Project effort estimation is part of the software development lifecycle. A realistic estimate of the effort in the initial phase of the project is necessarily better to allocate resources for the development of the project [3].According to [4], in the development phase, some artifacts are not yet consistent, allowing changes in the requirements and the effort estimation of the project, causing a great challenge for the project manager, since from the estimates, the changes can be accepted or rejected.

Without a realistic estimate, a software development project cannot be effectively managed. Today, using data mining techniques, in particular, machine learning (ML) algorithm, is used for effort prediction minimizing uncertainties [5].Several machine learning models have been proposed to predict software development project effort, such as Artificial Neural Networks, Decision Trees, Classification and Regression Tree, Bayesian Networks, Support Vector Machine, Genetic Programming, K-neighbors more nearby, and Extreme Machine Learning[6].
In the context of ML, software effort estimation is a regression problem. The regression algorithms are an equation that aims to estimate the value of a variable (y) based on one or more independent variables (x), given a history of correlation between the two variables. The function seeks to establish a linear function between X and Y that can determine the value of variable X according to the value of variable Y[7]. For the construction of regression models to estimate software development project effort, resources obtained before or during software development are used as input variables for the model[8].

In ML there is still the ensemble technique. The ensemble is a set of models formed by more than one ML algorithm. This technique has gained considerable popularity due to its good generalization performances. The ensemble generally results in better accuracy and is more stable than individual techniques, as they combine the results of its components to provide a single result. It is expected that with the creation of an ensemble if any of the models perform poorly, the system can reduce the error using many models [9].

In the area of effort estimation, the ensemble is not widely adopted. For example, in the work of [10] using Neural Network Ensemble, the models MLPNN Model, Ridge-MLPNN Ensemble Model, Lasso-MLPNN Ensemble Model, Bagging-MLPNN Ensemble Model, AdaBoost-MLPNN Ensemble Model are used in[11].

In this context, the major reasons for proposed this work are: (i) the ensemble makes the model more robust and stable, ensuring excellent carry out and through a set of two or more techniques [12], it can be performed. (ii) Using ensemble models to estimate the variables related to the effort estimation brings gain to this context and the various stakeholders in its applicability. They are tools that can be widely used, generating knowledge, serving as a basis for problem-solving and developing mechanisms to support the manager project.(iii) to reduce the gap in the academic literature[9][10] in the use of ML techniques used in the Bagging and Stacking set in the context of effort estimation. It is worth mentioning, in other regards, to obtain better performance in prediction[8] because the ensemble models create several linear models in different parts of the data set and then generalize them to get a more accurate prediction in the effort estimation for software design.

Therefore, we propose to create an Ensemble Regression model for estimating the effort of software projects. We use the ensemble bagging and stacking models in combination with the regression technique. In our experiments, we used the software effort data set available at and applied Bagging to the following predictors: Bagging with Linear Regression (B-LR), Bagging with Robust Regression (B-RR), Bagging with Ridge Regression (B-RI), Bagging with Lasso Regression (B-LA), Stacking with Ridge, Robusta, Lasso and Linear meta-predictor (ST-LR), Stacking with Linear, Ridge, Lasso and Robusta meta-predictor (ST-RR), Stacking with Linear, Robusta, Lasso and Ridge meta-predictor (ST-RI), Stacking with Linear, Robusta, Ridge, and Lasso meta-predictor (ST-LA).

The contributions of this paper are the ensemble models applied to the software project effort estimation field, as following: (i) use of the ensemble stacking and bagging methods for effort estimation, and (ii) comparison of the proposed models with models from the literature that use this dataset.

This paper is organized as follows. Section 2 presents related works. Section 3 presents the Background. Section 4 presents the methodology employed in this research. Section 5 presents the results of the experiments. Final considerations and future work are in section 6.

## 2. RELATED WORKS

In this section, we can observe research on the use of ML to estimate software development project effort.

In[13]the authors aim to improve the estimation of the software effort by incorporating direct mathematical principles and artificial neural network techniques. The process consists of transforming the problem of estimating the effort of the software into the problems of classification and functional approximation using a feed forward neural network. The results were systematically compared with previous related works, using only a few resources obtained, but they demonstrate that the proposed model produced satisfactory estimation accuracy based on the MMRE-Mean Magnitude of the Relative Error and PRED-Percentage Relative Error Deviation.

In [14]the authors proposed to apply a genetic algorithm to simultaneously select the optimal input feature subset and the parameters of a machine learning technique used for regression. The paper investigated three machine learning techniques: (i) Support Vector Regression (SVR), (ii) MLP Neural Networks, and (iii) Decision Trees (M5P). The genetic algorithm-based method showed a better performance, compared to the three machine learning techniques for software development effort estimation problems.

In[15]Gabrani and Saini conducted a comparative study of non-algorithmic techniques used for software development effort estimation. An empirical evaluation of five learning algorithms was carried out, namely: Fuzzy Learning based on Genetic Programming Grammar Operators (GFS-GPG-R), Symbolic Fuzzy-Valued Data Learning based on Genetic Programming Grammar Operators and Simulated Annealing (GFS-SAP- Sym-R), Symbolic Fuzzy Learning based on Genetic Programming Grammar Operators and Simulated Annealing (GFS-GSP-R), Ensemble Neural Network for Regression Problems (Ensemble-R), Fuzzy and Random Sets Based Modeling (FRSBM). Out of which the first three are variants of hybridization of genetic programming and fuzzy learning algorithms, the fourth one involves ensembling of neural networks, and the fifth one involves fuzzy random set based modeling. The proposed results are compared with other machine learning methods, such as MLP, SRV and ANFIS-Adaptive Neuro Fuzzy Inference Strategy. Based on the entire study, it is concluded that evolutionary algorithms give better results for the estimation of software effort compared to other machine learning methods. Of the five evolutionary learning algorithms that presented, the best result was the GFS- SAP-Sym-R.

The work conducted by Azzeh[16]presents a new approach to improve the accuracy of the effort estimate based on the use of the Optimized Tree Model. The bees algorithm was used to search for the optimal values of the Tree Model parameters to construct the software effort estimation model. As a reference, the results were compared to those obtained with gradual regression, case-based reasoning, and multilayer perceptron. The combination of the tree and bees model algorithm surpassed other well-known estimation methods.

In their work [17], Tierno and Nunercompared the data-driven Bayesian Networks with the regression of ordinary least squares with unique logarithmic transformation and with the mean and median baseline models. According to the author, BN has the potential for data-based predictions, but still need improvements to keep pace with more accurate data-based models.

In the work of[18], the authors evaluate automated ensembles of learning machines manages to improve software development effort estimation given by single ML and which of them would be more useful. In addition, the use of resource selection and regression trees (RTs) was analysed. Two personalized ways of combining sets and locations were investigated to provide additional information on how to improve software effort estimates. Bagging ensembles of RTs was among the best approaches for each data set. However, performance is significantly worse compared to the best approach for data sets.

In[19], the authors validated an automated genetic structure, carrying out sensitivity analyses in different genetic configurations to increase the forecast performance and optimize the processing time. The search space was represented by the combination of eight pre-processors, fifteen modeling techniques, and five attribute selectors. Through the elitism technique, the genetic structure selects the best combination of processing, attributes, and learning algorithm with the best correlation of coefficients. The metrics used for validation were: Spearman's rank correlation, MMRE - Mean of the Magnitude of Relative Error, MdMRE - Median of the Magnitude of Relative Error, MMAR - Mean of the Absolute Residuals, SA - Standardized Accuracy, and Pred25 - number of Predictions within % of the actual ones. They concluded that the study was able to improve some forecasting models based on the results of the best performance of the learning schemes, and that, according to the data set used for forecasting, the selection of an appropriate estimation technique directly impacts its performance.

In the work of[20], a study was carried out with four machine learning algorithms to create models for estimating software development effort. Artificial Neural Network (ANN), Support Vector Machines (SVM), K-star, and Linear Regression were evaluated using public data from software projects. The model that had the best performance was the SVM.

In[21], the authors proposed working with three machine learning models to increase the performance of the software effort estimation process, such as Multi-Layer Perceptron Neural Network (MLPNN), Probabilistic Neural Network (PNN), and Recurrent Neural Network (RNN). The result of his study suggests that the MLPNN model performed better compared to the other models, with 79% of successful estimates.

Therefore, given the presented scenario, this work differs from those previously shown since the ensemble models used and the respective composition of techniques proposed in this article were not applied to the dataset for effort estimation problems. The use of ensemble models combines more than one regression model, so there are specific models for each region, providing a more efficient estimation. Thus, these models enable better accuracy in estimating the final result of the effort estimation in software projects.

## 3. BACKGROUND

In this section, we present the notions related to ensemble techniques, Bagging (Bootstrap Aggregating), and Stacking, which are the prediction model applied in this research.

### 3.1 Bagging (Bootstrap Aggregating)

The *Bootstrap Aggregation* method, also called bagging, was one of the first predictor combination methods proposed by Breiman in 1996 [22], which generates a set of data by bootstrap sampling of the original data.

Bagging generates several training sets with data replacement and then builds a model for each set using the same machine learning algorithm[8].

In bagging, in a regression problem, the application can be described as a training sample $D_t$ = $(x_1, y_1), ..., (x_n, y_n)$, whose instances are independent from a probability distribution $P(x, y)$. Hence, bagging combines the prediction of a collection of regressors, in which each regressor is constructed by applying a fixed learning algorithm to a different bootstrap sample from the original $D_t$ training data. In equation 1 is described the representation of bagging, which the forecast on the set is the average of the individual forecasts of the generated $M$ regressors.

$$fbag(x) = \frac{1}{M} \sum_{i=i}^{M} \hat{f}_i(x) \tag{1}$$

where, *fbag(x)* is a combined forecast model for time *x*, $M$ is the number of components in the model and $\hat{f}_i(x)$ is an output of the base component.

### 3.2 Stacking

Stacking is a technique used to combine several models. The idea is to gather the advantages of different techniques, minimize the error rate of the models, and create a meta-predictor that combines the outputs of different models [23].

One of the ways to build a stacking is to collect the outputs of each model that makes up a set to form a new set of data. As [24], there are two learning levels 0 and 1. Level-0 are models trained and tested in independent cross-validation examples from the original data set. The output of this model and the original input data are used as input for level-1, called generalized, that is, the meta predictor. In this way, level-1 is developed using the results of level-0 generalizers.

Through the *M* set of predictors (linear or non-linear), instead of selecting just a single model from this set, a more accurate predictor can be obtained by combining the *M* predictors. The idea is that the level-1 data (result formed by each *M* predictor) has more information and can be used to build good combinations of predictors. Equation 2 shows the formulation of the stacking function,

$$fstack(x) = \sum_{j=1}^{K} \left[ f(x_j) - \sum_{i=1}^{M} \propto_i * \hat{f}_i(x_j) \right] \tag{2}$$

where, *fstack(x)* is the combined model prediction for *x*, *K* is the combined training data size, *M* is the number of model regressors, *α* is one coefficient that minimizes the error, and $\hat{f}_i(x_j)$ is the prediction given by the *i-th* built regressor.

## 4. METHODOLOGY USED IN THE IMPLEMENTATION

In this section, we will present the methodology based on[25] that was used in this work. The methodology(Figure 1) consists of three phases: data preparation, modeling and experimental evaluation.

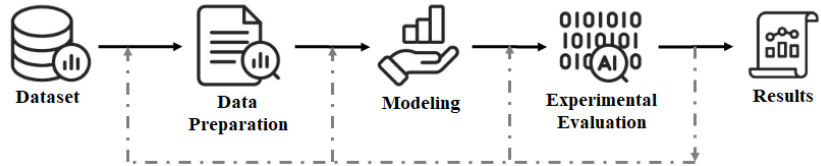

Figure 1. Methodology

### 4.1 Data Preparation

Knowing the type of data is fundamental for choosing the appropriate methods to be used. The dataset analysed and the summary is available at [26], which has quantitative and qualitative information about projects.

The independent variables of the models are "N&C" New and Changed as well as "R" (Reused) and all of them are considered as physical lines of code (LOC). N&C constituted of added and modified code. The joined code is the LOC written during the current programming process, while the modified code is the LOC changed in the base program when modifying a previously developed program.

The correlation or correlation coefficient measures the tendency for two variables to change depending on their relationship. Pearson's correlation coefficient produces a result between -1 and 1. A result of -1 means that there is a perfect negative correlation between the two values. In contrast, a result of 1 means that there is a definite positive correlation between the two variables. Thus, Figure 2 (a) shows a high correlation between N&C and AE (effort) and Figure 2 (b) exhibitions a low correlation between R-used and AE (effort). Therefore, an increase in the N&C variable increases the AE-effort, indicating cause and effect relationships.

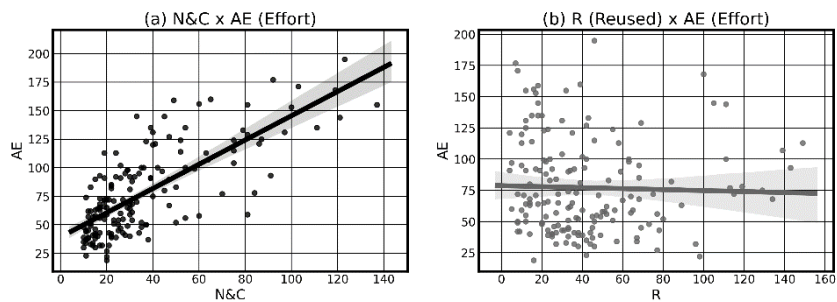

Figure 2. Correlation between N&C and AE-effort (a) and Correlation between R-reused and AE-effort (b)

Figure 3 illustrates the histogram of the dependent variable (Ae-effort) relative to effort, which is measured in minutes. The histogram shows a slight tendency to form a normal distribution, where the highest concentration of data is around the mean and the frequency near to limits.

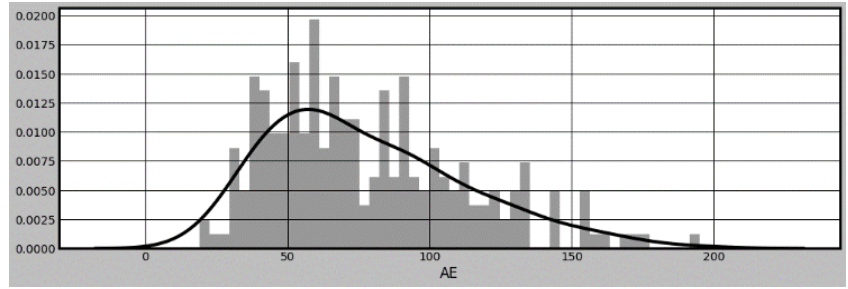

Figure 3. Effort Value Distribution

Summary statistics for the numerical variables from the database are given in Table 1. In addition, there is a significant difference between Mean and Median to numerical variables, indicate the existence of outliers values.

Table 1. Summary Statistics for numerical variables.

| Variable | Description | Mean | Stdev | Median |
|---|---|---|---|---|
| N&C | New and Changed code | 35.56 | 26.60 | 27.00 |
| R | Reused code | 41.82 | 30.86 | 34.00 |
| AE | Actual Effort (minutes) | 77.07 | 37.81 | 67.00 |

To solve the outlier problem, we normalized the data between 0 and 1. In normalization, the Max-Min method was utilized to employ the maximum and minimum values of the variable in question and its standard deviation to normalize the data on a scale uniform.

### 4.2 Modeling

We used the bagging [22]method in the modeling phase to generate a bootstrap sample set of the original data. This dataset will generate a set of models using a simple learning algorithm by combining their means. It's according to *Equation (1)* and the ensemble stacking *Equation (2)* described in Section 3. Thus, the eight models were proposed:

- ProposedModel1: ensemble Bagging with linear regression (here called *B-LR*).
- ProposedModel2: ensemble Bagging with robust regression (here called *B-RR)*.
- ProposedModel3: ensemble Bagging with ridge regression (here called *B-RI*).
- ProposedModel4: ensemble Bagging with lasso regression (here called *B-LA*).
- ProposedModel5: ensemble Stacking with robust regression, ridge regression, lasso regression, and meta-predict (linear regression) (here called ST-LR).
- ProposedModel6: ensemble Stacking with linear regression, ridge regression, lasso regression, and meta-predict (robust regression) (here called ST-RR).
- ProposedModel7:ensemble Stacking with robust regression, linear regression, lasso regression, and meta-predict (ridge regression) (here called ST-RI).
- ProposedModel8: ensemble Stacking with lasso regression, linear regression, ridge regression, and meta-predict (lasso regression) (here called ST-LA).

The eight models proposed were compared with the literature models, which used the same dataset for estimating effort in software development. The models in the literature are:

- Linear Regression[26].
- ELM (Extreme Learning Machine)[27].

## 4.3 Experimental Evaluation

This section depicts the carry out the measurements used in this research. There are diverse metrics used in the literature to evaluate the accuracy of prediction methods in software project effort estimation[28]. According to[28]an evaluation metric that does not have a good precision for the prediction of software effort estimation is MMRE.

To qualify the performance of the model, which is compared with each technique, the performance index used in this work is the Mean Absolute Residual (MAR), as denoted in *Equation 3*.

$$MAR = \frac{\sum_1^n |y_i - \hat{y}_i|}{n} \quad (3)$$

$y_i$ is the ith value of the variable being predicted, $\hat{y}_i$ its estimate, $y_i - \hat{y}_i$ the ith residual and n the number of cases in the dataset.

The Relative Gain (RG)[9]is another form of measurement is to analyse the proposed models about related works. The aim of the RG is measuring the gain about minimizing any prediction error, and this value shown in percentage. The RG presents in *Equation 4*.

$$RG = 100 * \left(\frac{Erro_a - Erro_b}{Erro_a}\right) \quad (4)$$

With the sample of 1000 iterations, we calculate the standard deviation (SD) of the error. Besides, we performed statistical tests, such as the Kolmogorov-Smirnov and Wilcoxon tests. We also use boxplots and relative p-value also to evaluate the performance of the models.

## 5. RESULTS AND DISCUSSION

In this section, we present the results obtained in the experiments. It was composed by the ensemble bagging and stacking methods. Besides, the ensemble models were created with parametric techniques, and performed a Monte Carlo simulation with 1000 iterations on the literature and proposed models. Algorithm 1 presents the pseudocode for experimental evaluation.

| Algorithm1: Pseudo-code of the experiment execution |
|---|
| 1 **Input:** Use the dataset |
| 2     **Set:** Number Simulation (MC) = 1000 |
| 3     **For** all i = 1 to MC do it: |
| 4         **Shuffle** dataset Training (70%) and Test (30%) |
| 5         **Apply:** ensemble models (B-LR,B-RR,B-RI,B-LA,ST-LR,ST-RR,ST-RI,ST-LA) to training set |
| 6         **Calculate**: the (MAR) of the models |
| 7     **endfor** |
| 8     **Calculate** mean and standard deviation of the error (MAR), *Equation 3* |

Table 2 shows the error (Equation 3) of the mean and standard deviation of the ensemble proposed models. We observed in the results that the averages of the models with lasso regression are smaller (B-LA, ST-LA), indicating that these two models had the best performance.

Table 2. Results Means and Standard Deviation.

| Techniques | Mean (standard deviation) |
|---|---|
| B-LR | 23.8934 ($3.6888*10^{-1}$) |
| B-RR | 23.4201 ($3.9832*10^{-1}$) |
| B-RI | 21.8855 ($2.4547*10^{-1}$) |
| B-LA | 21.5707 ($1.2946*10^{-1}$) |
| ST-LR | 23.5129 ($3.3108*10^{-14}$) |
| ST-RR | 23.2779 ($1.6313*10^{-12}$) |
| ST-RI | 22.6649 ($1.5244*10^{-13}$) |
| ST-LA | 21.5507 ($5.0822*10^{-14}$) |

Figure 4 shows the boxplot graph of the ensemble models. In addition, we demonstrate that there is an outlier in all proposed bagging models (B-LR, B-RR, R-RI and B-LA), and the variance is little and similar in proposed stacking models (ST-LR,SR-RR,ST-RI and ST-LA). We conclude that the proposed stacking models are less sensitive to outliers.

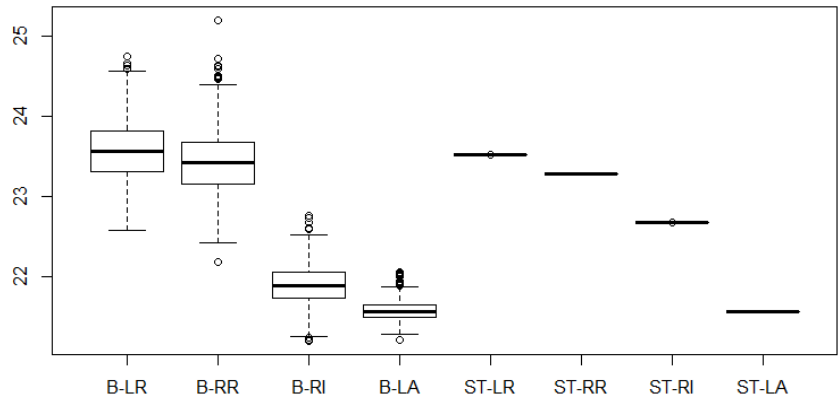

Figure 4. Boxplot Model

The Kolmogorov Smirnov normality test [29]was used. All datasets do not carry on a normal distribution. Thus, the Wilcoxon hypothesis test[30]with a significance of 5% was quantified. The alternative hypothesis is that the B-LA model had smaller errors (H1), and the null hypothesis is that the models have the same errors (H0).Equation 5.

$$\begin{cases} H_0 : \mu_1 = \mu_2 \\ H_1 : \mu_1 < \mu_2 \end{cases} \quad (5)$$

Table 3shows the result of the *p-value* for the Wilcoxon tests. According to the analyses, the ST-LA model does not present statistical evidence of minor errors than the B-LA. Thus, proposed stacking ensemble models using parametric techniques have promising results for the problem of the effort estimation in software development.

Table 3. Results *p-value*.

| Techniques | *p-value* |
|---|---|
| ST-LA vs B-LR | $2.2*10^{-16}$ |
| ST-LA vs B-RR | $2.2*10^{-16}$ |
| ST-LA vs B-RI | $2.2*10^{-16}$ |
| ST-LA vs B-LA | 0.19144 |
| ST-LA vs ST-LR | $2.2*10^{-16}$ |
| ST-LA vs ST-RR | $2.2*10^{-16}$ |
| ST-LA vs ST-RI | $2.2*10^{-16}$ |

Table 4shows the comparison with other studies of effort estimation in software development. We compare the datasets and types of ensemble used in the articles.

Table 4. Related Work Comparation.

| Authors | Dataset | Evaluation | Techniques |
|---|---|---|---|
| Kultur et al.[31] | NASA, NASA 93, USC, SDR, Desharnais | MMRE | ENNA, ENN, NN |
| Pai et al.[10] | 163 projects from a leading CMMI level 5 | MRE | Neural Network, Ensemble |
| Elish and Helmy [32] | Albrecht, Miyazaki, Maxwell, COCOMO, Desharnais | MMRE | SVR, MLP, ANFIS |
| Kocaguneli et al.[33] | COCOMO81, NASA93, Desharnais, SDR | MRE, MMRE | CART |
| Shukla et al.[11] | 81 software projects from a Canadian software company (PROMISE) | $R^2$ | MLPNN Model, Ridge-MLPNN Ensemble Model, Lasso-MLPNN Ensemble Mode, Bagging-MLPNN Ensemble Model, AdaBoost-MLPNN Ensemble Model |
| Abnane et al.[34] | Albrecht, COCOMO81, Kemerer, Desharnais, ISBSG, Miyazaki | MAE | E-KNNI, GS-KNNI, UC-KNNI |

The eight proposed ensemble models are different from the related works show in Table 4. We take care to use parametric methods for building the ensemble models. Also, we used another dataset for effort estimation in software development.

Some articles use the study dataset. But, the techniques used are linear regression and ELM (Extreme Learning Machine) with 2 and 5 n_hidden. We compared the eight proposed models with the developed ones.Equation6[26]presents the coefficient of linear regression used during the comparison, in which N&C and R (Reused) is previously described in Section 4.

$$Effort = 44.713 + (1.08 * N\&C) - (0.145 * R) \qquad (6)$$

We observed the error (Equation 3) of the mean and standard deviation of the literature models in Table 5, and all errors were higher than those obtained by the proposed model's majority. Therefore, there is a big difference in the errors obtained with other techniques in this same dataset in comparison with proposed ensemble models.

Table 5. Comparison of Literature results (Error).

| Techniques (Ref.) | *Error (SD)* |
|---|---|
| Linear Regression ([26]) | 48.5674 (1.3587*10$^{-13}$) |
| ELM with 2 n_hidden ([27]) | 23.8934 (3.9770*10$^{0}$) |
| ELM with 5 n_hidden ([27]) | 24.228 (2.0757*10$^{0}$) |

Figure 5shows the boxplot graph of the ELM models. Analyzing it is noted a significant presence of ELM outlier with 2 n_hidden. However, the model with 5 n_hidden still presents an average with larger errors, since most of the error values are above the median of the other ELM model.

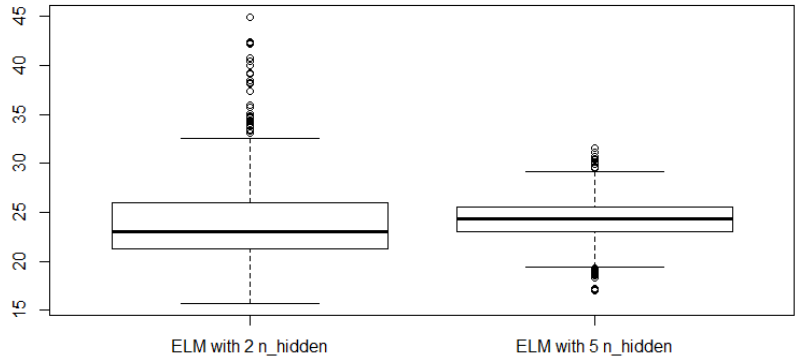

Figure 5. Boxplot Model

Table 6 presents the RG (Equation 4)of proposed models about the related works. We can verify that the obtained gain was very significant. We demonstrated that proposed models are more efficient than other literature models. We showed that the proposed prediction models fit well with ensemble bagging and stacking methods, considering the resultant effect of the increase in accuracy, reduced error rate as well as improvement in predictive efficiency. It can ratify the mean values obtained in Table 2.

Table 6. Result of RG.

| Techniques | Linear Regression | ELM with 2 n_hidden | ELM with 5 n_hidden |
|---|---|---|---|
| B-LR | 50.80% | 0% | 1.38% |
| B-RR | 51.77% | 1.98% | 3.33% |
| B-RI | 54.93% | 8.40% | 9.66% |
| B-LA | 55.58% | 9.72% | 10.96% |
| ST-LR | 51.58% | 1.59% | 2.95% |
| ST-RR | 52.07% | 2.57% | 3.92% |
| ST-RI | 53.33% | 5.14% | 6.45% |
| ST-LA | 55.62% | 9.80% | 11.05% |

The main contribution of this work is that the eight proposed ensembles models present better accuracy for this study's research problem. Also, use bagging and stacking with parametric techniques formation the models. The novel of this research and technical study is the application of ensembles models in the dataset. Thus, the accuracy of the estimation of software development enables companies to know the amount of effort required to develop this application on time and within budget, before implementing an application. Also, to estimate effort, it is generally necessary to know previous similar projects that have already been developed by the company and understand the project variables that may affect effort prediction in software development.

## 6. THREATS TO VALIDITY

According to[35] describe threats to validity, it is clear that all the limitations presented are categorized as external validity. External validity can aim when studying is relevant to others considering the sample quantity.

Ensemble Regression models were used during various stages of the present research. Considering the random nature of these models, the results obtained from every implementation might appear a little different from one another. MAR and RG used in the present article are biased. They have only chosen herein because they were found most frequently employed in prior research.

Order to estimation training and test set, all of the data were randomly assigned to training and test sets in 70 % to 30 % ratio, respectively. The random assignment of the data can have a considerable influence on the model results. However, considering that all the models are run single datasets, there will be made not much of an effect on the overall work since the objective has been to compare the performance of various ensemble models on the dataset applied.

We have some limitations regarding the size of the data set, as well as the number of attributes used to estimate the effort in software projects. The availability of data from software projects is another limitation, as the availability of data is not frequent, causing difficulties in forecasting with a reduced amount of data. Therefore, the number of instances in the data set must be more significant.

According to the results, satisfactory outputs were obtained due to the useful findings (lower prediction errors). However, it can be seen that the eight ensemble models proposed herein have had better performance concerning the literature models.

## 7. CONCLUSIONS AND FURTHER WORK

Accurate estimation of software project effort at an early stage in the development process is an important challenge for the software engineering community. In this direction, this research lavished attention on the issues related to software effort estimation using ensemble models.
Therefore, this work presents models for effort estimation of software projects to serve as a decision support tool for project managers in the process of specification, development, maintenance, and creation of software, aiming at the productivity and quality of the projects.
According to the related works, many articles used the Mean Magnitude of Relative Error (MMRE) to assess the accuracy of the forecasting methods in estimating the effort of the software project. However, this accuracy is not a reliable indicator of forecast evaluation in the estimation of the software project effort. Therefore, in this article, we use MAR as an error estimate.

In our simulations, we used a dataset of software projects similar to our reality, 163 small programs developed by 53 programmers, and validation 68 programs developed by another group, integrated by 21 programmers.

We also conducted experiments to compare the dataset and techniques to the related works. We used eight ensemble regression models based on bagging and stacking methods. The main contributions of this paper are:

1. Comparison between ensemble regression models in the context of effort estimation. It could increase efficiency, reduced error rate, and increase in accuracy predictive;
2. We showed through the experiments that the proposed models got better results compared with literature models;
3. The proposed ensemble regression models (B-LR, B-RR, B-RI, B-LA, ST-LR, ST-RR, ST-RI, and ST-LA) allow identifying the estimation of the effort the form efficient.
4. The accuracy in estimating effort enables project managers to determine the duration, staffing, and cost required for software development.

It is concluded that using machine learning techniques to estimate software development efforts enhances the projects to have more chances of success. Therefore, several investigations of other regression problems can be defined as future works of this study, including Boosting and Random Forest. Also, other data sets can be used for the experience and training of the model to compare the accuracy results.

## ACKNOWLEDGMENTS

This study was financed in part by the Coordenação de Aperfeiçoamento de Pessoal de Nível Superior - Brasil (CAPES) - Finance Code 001, FACEPE, and CNPq.

## REFERENCES

[1] R. S. Pressman, Software Engineering A Practitioner's Approach 8th Edition. 2016.

[2] B. Peischl, M. Nica, M. Zanker, and W. Schmid, "Recommending effort estimation methods for software project management," Proc. - 2009 IEEE/WIC/ACM Int. Conf. Web Intell. Intell. Agent Technol. - Work. WI-IAT Work. 2009, vol. 3, pp. 77–80, 2009.

[3] W. Han, H. Jiang, X. Zhang, and W. Li, "A Neural Network Based Algorithms for Project Duration Prediction," Proc. - 7th Int. Conf. Control Autom. CA 2014, pp. 60–63, 2014.

[4] J. Shah, N. Kama, N. A. A Bakar, and Z. Bhutto, "Software Requirement Change Effort Estimation Model Prototype Tool for Software Development Phase," Int. J. Softw. Eng. Appl., vol. 10, no. 03, pp. 09–19, 2019.

[5] P. Pospieszny, B. Czarnacka-Chrobot, and A. Kobylinski, "An effective approach for software project effort and duration estimation with machine learning algorithms," J. Syst. Softw., vol. 137, pp. 184–196, 2018.

[6] A. García-Floriano, C. López-Martín, C. Yáñez-Márquez, and A. Abran, "Support Vector Regression for Predicting Software Enhancement Effort," Inf. Softw. Technol., vol. 97, pp. 99–109, 2018.

[7] W. O. Bussab and P. A. Morettin, Estatística Básica, 9th ed. Pinheiros: Saraiva, 2017.

[8] P. L. Braga, A. L. I. Oliveira, G. H. T. Ribeiro, and S. R. L. Meira, "Bagging predictors for estimation of software project effort," IEEE Int. Conf. Neural Networks - Conf. Proc., no. October 2016, pp. 1595–1600, 2007.

[9] P. M. Da Silva, M. N. C. A. Lima, W. L. Soares, I. R. R. Silva, R. A. De Fagundes, and F. F. De Souza, "Ensemble regression models applied to dropout in higher education," Proc. - 2019 Brazilian Conf. Intell. Syst. BRACIS 2019, pp. 120–125, 2019.

[10] D. R. Pai, K. S. McFall, and G. H. Subramanian, "Software effort estimation using a neural network ensemble," J. Comput. Inf. Syst., vol. 53, no. 4, pp. 49–58, 2013.

[11] S. Shukla, S. Kumar, and P. R. Bal, "Analyzing effect of ensemble models on multi-layer perceptron network for software effort estimation," Proc. - 2019 IEEE World Congr. Serv. Serv. 2019, vol. 2642–939X, pp. 386–387, 2019.

[12] N. García-Pedrajas, C. Hervás-Martínez, and D. Ortiz-Boyer, "Cooperative coevolution of artificial neural network ensembles for pattern classification," IEEE Trans. Evol. Comput., vol. 9, no. 3, pp. 271–302, 2005.

[13] P. Jodpimai, P. Sophatsathit, and C. Lursinsap, "Estimating software effort with minimum features using neural functional approximation," Proc. - 2010 10th Int. Conf. Comput. Sci. Its Appl. ICCSA 2010, pp. 266–273, 2010.

[14] A. L. I. Oliveira, P. L. Braga, R. M. F. L. Lima, and M. L. Cornélio, "GA-based method for feature selection and parameters optimization for machine learning regression applied to software effort estimation," Inf. Softw. Technol., vol. 52, pp. 1155–1166, 2010.

[15] G. Gabrani and N. Saini, "Effort estimation models using evolutionary learning algorithms for software development," 2016 Symp. Colossal Data Anal. Networking, CDAN 2016, 2016.

[16] M. Azzeh, "Software Effort Estimation Based on Optimized Model Tree Mohammad," Proc. 7th Int. Conf. Predict. Model. Softw. Eng. PROMISE 2011, pp. 20–21, 2011.

[17] I. A. P. Tierno and D. J. Nunes, "An extended assessment of data-driven Bayesian Networks in software effort prediction," Proc. - 2013 27th Brazilian Symp. Softw. Eng. SBES 2013, pp. 157–166, 2013.

[18] L. L. Minku and X. Yao, "Ensembles and locality: Insight on improving software effort estimation," Inf. Softw. Technol., vol. 55, no. 8, pp. 1512–1528, 2013.

[19] J. Murillo-Morera, C. Quesada-López, C. Castro-Herrera, and M. Jenkins, A genetic algorithm based framework for software effort prediction, vol. 5, no. 1. Journal of Software Engineering Research and Development, 2017.

[20] M. Hammad and A. Alqaddoumi, "Features-level software effort estimation using machine learning algorithms," 2018 Int. Conf. Innov. Intell. Informatics, Comput. Technol. 3ICT 2018, pp. 1–3, 2018.

[21] S. Shukla and S. Kumar, "Applicability of Neural Network Based Models for Software Effort Estimation," Proc. - 2019 IEEE World Congr. Serv. Serv. 2019, vol. 2642–939X, pp. 339–342, 2019.

[22] L. Breiman, "Bagging Predictors," Mach. Learn., vol. 24, no. 421, pp. 123–140, 1996.

[23] A. A. Ghorbani and K. Owrangh, "Stacked generalization in neural networks: Generalization on statistically neutral problems," Proc. Int. Jt. Conf. Neural Networks, vol. 3, pp. 1715–1720, 2001.

[24] P. Kraipeerapun and S. Amornsamankul, "Using stacked generalization and complementary neural networks to predict Parkinson's disease," Proc. - Int. Conf. Nat. Comput., vol. 2016-Janua, pp. 1290–1294, 2016.

[25] U. Fayyad, G. Piatetsky-Shapiro, and P. Smyth, "From Data Mining to Knowledge Discovery in Databases," Am. Assoc. Artif. Intell., vol. 17, pp. 37–54, 1996.

[26] C. Lopez-Martin, "A fuzzy logic model for predicting the development effort of short scale programs based upon two independent variables," Appl. Soft Comput. J., vol. 11, no. 1, pp. 724–732, 2011.

[27] S. K. Pillai and M. K. Jeyakumar, "Extreme Learning Machine for Software Development Effort Estimation of Small Programs," Int. Conf. Circuit, Power Comput. Technol., pp. 1698–1703, 2014.

[28]  M. Shepperd and S. MacDonell, "Evaluating prediction systems in software project estimation," Inf. Softw. Technol., vol. 54, no. 8, pp. 820–827, 2012.

[29]  H. W. Lilliefors, "On the Kolmogorov-Smirnov test for normality with mean and variance unknown," J. Am. Stat. Assoc., vol. 62, no. 318, pp. 399–402, 1967.

[30]  F. Wilcoxon, "Individual Comparisons by Ranking Methods," Biometrics Bull., vol. 1, no. 6, pp. 80–83, 1945.

[31]  Y. Kultur, B. Turhan, and A. Bener, "Ensemble of neural networks with associative memory (ENNA) for estimating software development costs," Knowledge-Based Syst., vol. 22, no. 6, pp. 395–402, 2009.

[32]  M. O. Elish, T. Helmy, and M. I. Hussain, "Empirical study of homogeneous and heterogeneous ensemble models for software development effort estimation," Hindawi Math. Probl. Eng., vol. 2013, 2013.

[33]  E. Kocaguneli, T. Menzies, and J. W. Keung, "On the value of ensemble effort estimation," IEEE Trans. Softw. Eng., vol. 38, no. 6, pp. 1403–1416, 2012.

[34]  I. Abnane, M. Hosni, A. Idri, and A. Abran, "Analogy Software Effort Estimation Using Ensemble KNN Imputation," Proc. - 45th Euromicro Conf. Softw. Eng. Adv. Appl. SEAA 2019, no. 1, pp. 228–235, 2019.

[35]  P. Runeson and M. Höst, "Guidelines for conducting and reporting case study research in software engineering," Empir. Softw. Eng., vol. 14, no. 2, pp. 131–164, 2009.

International Journal of Software Engineering & Applications (IJSEA), Vol.11, No.3, May 2020

**AUTHORS**

Halcyon Carvalho is a Project Manager, currently a Master's Degree student in Computer Engineering from the University of Pernambuco, postgraduate in Project Management, and Graduated in Information Systems. Experience in project management for 8 years in the IT area. Experience in IT project management covering activities related to software development (Factory). I am currently a member of the PMO team of the TRF5-Tribunal Regional Federal da 5ª Região, responsible for implementing the PMO.

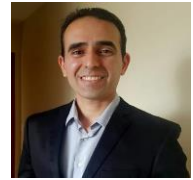

Marília Lima, has a degree in Information System from the University of Pernambuco (2017) and a master's degree in Computer Engineering from the University of Pernambuco (2019). Currently a Ph.D. student in Computer Engineering. Marília has experience in Computer Science, with emphasis on Computational Intelligence.

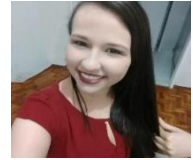

Wylliams Santos is an adjunct professor at the University of Pernambuco (UPE), where he leads the REACT Research Labs. Ph.D. in Computer Science (2018), Informatics Center (CIn) at Federal University of Pernambuco (UFPE), Brazil. MSc in Computer Science (2011), Informatics Center at Federal University of Pernambuco, Brazil. He undertook his sandwich PhD (2015-2016) research at the Department of Computer Science and Information Systems (CSIS) of the University of Limerick, Ireland and in collaboration with Lero - the Irish Software Research Centre. His research areas of interest includes management of software projects, agile software development and empirical software engineering.

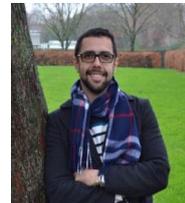

Roberta Fagundes, has a Post-Doctorate in Statistics (2015) from the Federal University of Pernambuco (UFPE). He also holds a doctorate (2013) and a master's degree (2006) in Computer Science from UFPE. Graduated in Telematics Technology (2002) from the Federal Center for Technological Education of Paraíba (CEFET-PB). He is currently an Adjunct Professor at the University of Pernambuco (2007) in the course of Information Systems and Computer Engineering at the University of Pernambuco (UPE). He is also vice-coordinator and professor of the Graduate Program in Computer Engineering (PPGEC), where there are Master's and Doctorate courses. Has interest in research in the area of Computer Science, with emphasis on Computational Intelligence.

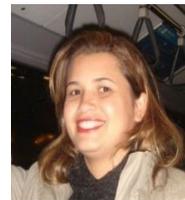